\documentclass[preprint,12pt]{elsarticle}

\usepackage{graphicx}

\usepackage{amssymb}
 
\usepackage{amsmath}
\usepackage{enumerate}

\journal{Journal of Biophotonics}

\begin{document}

\begin{frontmatter}
  
\title{{\em In situ} fiber-optical monitoring of cytosolic calcium in tissue explant cultures}

\author[iap]{Manuel Ryser}

\author[ana]{Lisa K\"{u}nzi}

\author[ana]{Marianne Geiser}

\author[iap]{Martin Frenz\corref{cor1}}

\author[iap]{Jaro Ri\v{c}ka}

\address[iap]{Institute of Applied Physics, University of Bern, Sidlerstrasse 5, 3012 Bern, Switzerland}

\address[ana]{Institute of Anatomy, University of Bern, Baltzerstrasse 2, 3000 Bern, Switzerland}

\cortext[cor1]{Correspondence should be addressed to M. F.~(email: martin.frenz@iap.unibe.ch).}

\begin{abstract}

We present a fluorescence-lifetime based method for monitoring cell and tissue activity {\em in situ}, during cell culturing and in the presence of a strong autofluorescence background. The miniature fiber-optic probes are easily incorporated in the tight space of a cell culture chamber or in an endoscope. As a first application we monitored the cytosolic calcium levels in porcine tracheal explant cultures using the Calcium Green-5N (CG5N) indicator. Despite the simplicity of the optical setup we are able to detect changes of calcium concentration as small as 2.5nM, with a monitoring time resolution of less than 1s. 

\end{abstract}


\end{frontmatter}





\section{Introduction}
\label{ch:introduction}
The topic of the present study, namely {\em in situ} and in-real-time monitoring of the physiological state of supported cell-cultures using a simple fiber-optic probe, is motivated by a specific project aimed at investigating responses of cultured respiratory tissue during exposure to nanoparticle aerosols and gases \cite{savi2008,Mertes2013}. Therein, a novel aerosol deposition chamber has been developed, designed for efficient deposition of nanoparticles on the cell cultures under conditions closely mimicking particle deposition conditions \textit{in vivo}. In-real time monitoring allows to correlate a stimulus of the cells to the acute physiological response and to investigate the direct effect of a stimulus on the physiological state of a cell.
The conditions in the exposure chamber pose the following constraints on the monitoring system: i) the monitoring probes must be small enough, so that they do not interfere with particle deposition, and ii) interference of the probing with the biological sample must be kept minimal. Both constraints suggest the use of optical meth\-ods, but constraint i) precludes \textit{in situ} on-line imaging of the cultures. Instead we use miniature probes that consist of bare optical fibers. 

There is a broad spectrum of fluorescence-based indicators available for almost any biochemical reaction, but our choice is somewhat restricted. Because of the nature of the observed sample (i.e. corrugated air-liquid interface of the cell culture) and because of the simplicity of the probing optics (i.e. bare optical fibers), we can not rely on a simple intensity measurement from a single fluorescence peak, but rather we must employ ratiometric or fluorescence-lifetime-based techniques. Therefore, we decided on the fluorescence lifetime that can be most sensitively and accurately measured with the time-correlated single photon counting (TCSPC) technique \cite{becker2005}.

\begin{figure}
	\centering
	\includegraphics[width=15cm]{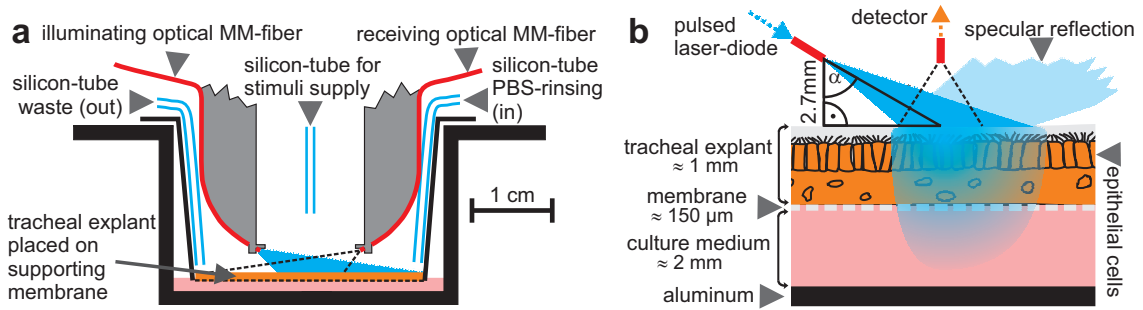}
	\caption{\label{fig:figure_1} Experimental setup. \textbf{(a)} The fiber-optic probe head fits into the cell culture insert around a central tube for stimuli application. Two additional silicon tubes allow for rinsing of the tissue. \textbf{(b)} Schematic of the sample irradiated by laser light delivered by a multimode optical fiber. The receiving fiber is placed at 90$^{\circ}$ (azimuthal) with respect to the illuminating fiber. }
\end{figure}

For the present feasibility study we chose to monitor the cytosolic calcium concentration in microdissected porcine tracheal tissue explants cultured on cell culture inserts at the air-liquid interface (ALI) within the afore mentioned exposure chamber with a supplemented high glucose dulbecco's modified eagle medium (DMEM, cp. supplementary material). 

Microdissected tracheal or bronchial explants mimic the \textit{in vivo} state of the respiratory epithelium best, as they contain the various highly differentiated cell types in their respective numbers and locations and produce their own surface lining layer liquid. On the other hand, they are particularly challenging from the point of view of fluorescence monitoring, because of the strong autofluorescence background due to the high elastin content of the connective tissue \cite{deyl1980} and riboflavin in the cell culture medium.

Calcium plays a fundamental role in the cellular physiology, since it acts as second messenger in signal transduction pathways of animal cells. Much research has been done to understand these complex processes and therefore many flu\-o\-res\-cent indicators are available to measure cytosolic calcium \cite{czarnik1992,Takahashi1999}. The Calcium Color Series (Calcium Green, Calcium Orange, Calcium Crimson) are suitable for lifetime-based sensing. According to the findings from Yoshiki et al. \cite{Yoshiki2005} CG5N appears to be best suited for our purpose.

The CG5N indicator exhibits the following properties: (i) it shows high sensitivity for lifetime-based measurements \cite{Yoshiki2005}; (ii) it exhibits low affinity to calcium ($K_d\approx14000$nM) \cite{Yoshiki2005}; (iii) the calcium-affinity is insensitive to variations in pH around the physiological state of 7.2 \cite{Eberhard1991}; (iv) the chelator is highly sensitive to Ca$^{2+}$ over Mg$^{2+}$ ($>10^5$) \cite{Tsien1980}; (v) it can be introduced into the cells by incubation as described in Section \ref{calciumgreen_loading}.
 
The low affinity of CG5N is on the one hand very favorable since it means that the buffering-effect on intracellular calcium is minimized and thus also disturbance of the observed biological system. On the other hand it makes the detection of fluorescence changes more challenging since the low affinity fluorescence indicators emit much lower fluorescence-intensity levels compared to high affinity indicators. This disadvantage however, is irrelevant when using the highly sensitive single photon counting technology.

\section{Methods and materials}

\subsection{Sensing optics and cell culture chamber}

Our setup is illustrated in Fig.~\ref{fig:figure_1}. The sensing optics consists of two 50$\mu$m core diameter multimode-fibers (QMMJ-3AFX-UVVIS-50/125-3-4, OZ Optics LTD). Both, the illuminating and the receiving fiber, were used without any front-end optics. The numerical aperture N.A. $\approx0.2$ corresponds to the half angle $\theta\approx11.5^\circ$ of the illuminating cone. The fibers are bent and aligned to point at the same spot on the sample surface. It is crucial that the bending radius is not too small, otherwise the loss of guided light increases significantly. The illuminating light cone impinges onto the sample at an angle $\alpha\approx62^{\circ}$ to the surface normal. About 97\% of the light is transmitted through the air-tissue boundary, while about 3\% is reflected specular. To minimize the contribution of back-reflexion to the measured signal, the receiving fiber is oriented at 90$^{\circ}$ azimuthal with respect to the illuminating fiber.

The sample holder is placed in a self-built climatized cell exposure chamber heated and humidified by water circulation from a thermostatted water bath (RC6, MGW Lauda). Temperature and humidity of the air inside the chamber were monitored with a combined temperature and humidity sensor (SHT11, Sensirion AG) and the temperature of the cultured tissue was measured with a calibrated temperature sensor (Pt100). A measurement over 4.5 hours was performed to characterize the stability of the temperature and humidity inside the chamber: tissue temperature was stated to be 36.96$\pm$0.01$^\circ$C, air temperature 36.90$\pm$0.08$^\circ$C and relative humidity 92.7$\pm$0.8 r.H. 

\subsection{Data acquisition with Time Correlated Single Photon Counting technique}
The illuminating fiber guides the light of a pulsed laser diode (LDH-P-C-470B, PicoQuant GmbH, P$_{average}$=1.5mW @ 20,0MHz repetition frequency, $\lambda$=468nm, pulsewidth 63ps FWHM). The light collected by the receiving fiber is delivered to a single photon detector module (id100-50-STD, id Quantique SA). An optical long-pass interference filter in front of the detector ($>$500nm, XF3092/22, Omega Optical) blocks the directly scattered laser light. The photon timing module (SPC-530, Becker\&Hickl GmbH) measures the time lag between a detected flu\-o\-res\-cence photon and the laser pulse that generated it, i.e. the TCSPC system was triggered in the 'reverse-start-stop' mode \cite{becker2005}. 
The instrumental response of the setup, including the fiber optics, was measured at the excitation wavelength, with a single scattering and highly absorbing sample in place of the tissue. Thereby we neglect the well known dependence of the detector response on wavelength \cite{Ricka1981}, since this effect is compensated by our calibration based data evaluation. 

For time resolved measurements of the changes of calcium concentration we record typically 2000 consecutive timing histograms. The sensitivity of single photon timing and the available laser power allow us recording the calcium concentration with a sampling frequency of several hertz.

\subsection{Data evaluation procedure}
\label{sec:data_evaluation}
The observed samples consist of a relatively thick layer of tissue, a supporting PET-membrane and the culture medium (Fig.~\ref{fig:figure_1}b). Each layer contributes to the optical response to exciting light. Upon optimizing the excitation wavelength and the filters in the observation channel one still finds that the optical response $I(t)$ to short laser pulse excitation consists of three contributions: 
\begin{equation}
I(t)=F^I(t)+F^A(t)+S(t).
\label{eq:superpos_afl_stref0}
\end{equation}
Of interest is the nanosecond response of the indicator dye $F^I(t)$, but there is also auto-flu\-o\-res\-cence from the tissue and culture medium. We include these contributions in $F^A(t)$.
Finally, $S(t)$ represents the residual stray light due to scattering and reflections at optical interfaces. The stray light contributes only in the initial rise of the signal and its effect is expected to be negligible. However, the contribution $F^A(t)$ is rather strong (cp. Fig.~\ref{fig:figure_2}). The only way to recover the desired fluorescence response of the CG5N indicator molecules $F^I(t)$ is to determine accurately the time course autofluorescence decay $F^A(t)$, and include this information in the evaluation of $I(t)$. 
The procedure requires suitable models for both $F^I(t)$ and $F^A(t)$. In order to justify neglecting the stray light, we also need a model for $S(t)$.

\subsection{Fluorescence indicator model and calibration} 
CG5N features two forms of the probe $P$, namely a calcium bound $P_b$ and a free $P_f$ form, each characterized by a unique flu\-o\-res\-cence decay time. The fluorescence intensity decay in response to $\delta$-pulse excitation is bi-exponential:
\begin{equation}
	F^I(t)=A_{b}\,\exp\left(-t/\tau_{b}\right) + A_f \,\exp\left(-t/\tau_{f}\right).
	\label{eq:bi-exponential-impulse-response}
\end{equation}
Here $\tau_{b}$ and $\tau_{f}$ denote the flu\-o\-res\-cence decay time constants of the bound and free indicator dye molecules.
By fitting this model function to a recorded fluorescence decay signal one obtains the amplitudes $A_b$, $A_f$ and the life times $\tau_b$, $\tau_f$ from which the integral contributions $i_b=A_b\tau_b$ and $i_f=A_f\tau_f$ to the fluorescence signal are calculated. 
The integral contributions $i_n$ ($n$ = $b$ or $f$) are related to the concentration $\left[P_n\right]$ of a species by
\begin{equation}
i_n=C\,\langle \sigma_n \rangle \,q_n\,\left[P_n\right],
\label{eq:fractional_intensity_concentration}
\end{equation}
where $\langle \sigma_n \rangle$ denotes the orientational averaged absorption cross section and $q_n$ the quantum yield of the species. The common pre-factor $C$ accounts for the excitation intensity, beam profiles and angular apertures of excitation and observation, filter transmissions and detection efficiency. The concentrations $[P_f]$ and $[P_b]$ are related through
\begin{equation}
	K_d=\frac{\left[\mathrm{Ca}^{2+}\right]_{free} \times \left[P_f\right]}{\left[ P_b\right]},
	\label{eq:analyte_probe_equilibrium}
\end{equation}
where $K_d$ is the dissociation constant of the reaction $[P_b]\overset{K_d}{\rightleftharpoons} \left[\mathrm{Ca}^{2+}\right]_{free} + \left[P_f\right]$. 
Thus, we may write
\begin{equation}
\left[\mathrm{Ca}^{2+}\right]_{free}=K_d\,R\cdot\frac{A_{b}}{A_f }, \mbox{\hspace{.5cm}where\hspace{.5cm}} R=\frac{\langle\sigma_f \rangle \, q_f \, \tau_{b}}{\langle\sigma_{b} \rangle \, q_b\, \tau_f}.
\label{eq:calibration_equation}
\end{equation}
 An estimate of the product $K_d R$ can be obtained performing a series of calibration measurements with calcium buffers.

\subsection{Epithelial tissue autofluorescence and scattering model}
There are many intrinsic fluorophores incorporated in biological tissues and cell culture medium, each contributing its part to the auto-flu\-o\-res\-cence signal. Correspondingly, the time course of the auto-fluorescence decay can be expressed as the integral over a broad distribution $\hat{F}^A\left(k\right)$ of relaxation rates:
\begin{equation}
F^A\left(t\right)=\displaystyle \int_{0}^{\infty}{\hat{F}^A\left(k\right)\,e^{-kt}dk}.
\end{equation}
A typical decay curve resulting thereof is shown in Fig.~\ref{fig:figure_2}b.
In the present context we are not particularly interested in the molecular origins of the distribution $\hat{F}^A\left(k\right)$, but we need a mathematical function that allows us to model the observed decay accurately, yet with a minimum of parameters. As shown in Sections \ref{sec:epithelial_autofluorescence_decay}-\ref{sec:inf_of_temp_and_hum} these requirements are well fulfilled \cite{Siegel2003,BerberanSantos2007} by the so-called \textit{Stretched Exponential}, also known as the \textit{Kohlrausch-Williams-Watts} \cite{Jurlewicz1999,BerberanSantos2005} relaxation function:
\begin{equation}
F^A(t)=A_{s}\cdot\exp\left[-\left(t/\tau_s\right)^{1/h_s}\right],
\label{eq:stref}
\end{equation}
where $h_s\geq 1$ is the heterogeneity parameter ($h_s=1\rightarrow$ homogeneous) and $\tau_s$ is the mean decay time.

The diffusely backscattered laser impulses are, considering typical scattering properties for a typical biological tissue, temporally broadened by few 10ps (FWHM) \cite{Jacques1989,Patterson1989}. Thus, reflections and diffusely backscattered laser impulses can be approximated by a sum of $N$ Dirac delta function $\delta\left(t-t_{ri}\right)$:
\begin{equation}
S(t)=\sum_{i=1}^{N}{ A_{ri}\,\delta\left(t-t_{ri}\right)}.
\label{eq:model_straylight}
\end{equation}
Each laser impulse reflection, respectively each backscattered laser impulse, is characterized by an amplitude $A_{ri}$ and a temporal shift $t_{ri}$.


\subsection{Methodological approach}

The methodological approach is illustrated in Fig.~\ref{fig:figure_2}. A first step was to characterize the time dependent nanosecond autofluorescence response of the epithelial cells and their surrounding inside the cell exposure chamber (Fig.~\ref{fig:figure_2}b) in terms of the model function $F^A(t)$  (Eq.~\ref{eq:stref}). After the CG5N loading procedure (cp. Section \ref{calciumgreen_loading}) the cell cultures were placed back into the exposure chamber (cp. Fig.~\ref{fig:figure_1}b). The optical nanosecond response $I\left(t\right)$ (Fig.~\ref{fig:figure_2}a) following laser pulse irradiation contains a superposition (Eq.~\ref{eq:superpos_afl_stref0}) of the bi-exponential fluorescence decay from the CG5N indicator $F^I(t)$ (Fig.~\ref{fig:figure_2}c and Eq.~\ref{eq:bi-exponential-impulse-response}), autofluorescence from the biological tissue, cell culture medium and surrounding $F^A(t)$ (Fig.~\ref{fig:figure_2}b) and stray light $S(t)$ (Eq.~\ref{eq:model_straylight}). These signal contributions cannot be separated by spectral filtering because their spectra overlap (Fig.~\ref{fig:figure_2}d). Nevertheless they can be separated from each other by their temporal signature and the amplitude ratio $A_b/A_f$ (Eq.~\ref{eq:bi-exponential-impulse-response}) of the bound and free indicator dye molecules can be determined. By appropriate calibration this ratio can be correlated to calcium concentration $\left[\mathrm{Ca}^{2+}\right]_{free}$ (Eq.~\ref{eq:calibration_equation}). The numerical procedures are described in the supplementary material.


\subsection{Calibration protocol}
The calibration, i.e. the determination of the product $K_d\, R$ (cp. Eq.~\ref{eq:calibration_equation}), is achieved by re\-cording the response of the indicator dye to varying free calcium concentrations. Therefore we employed a commercially available EGTA-based calibration buffer kit (Molecular Probes, C-3009) adjusted to the composition of the intracellular solution, i.e., to a pH of 7.2 and an i\-on\-ic strength of 100mM KCl. With the EGTA buffer, it is crucial that the calibration solution is kept at a constant pH. At 37$^\circ$C and pH 7.2 the error in free calcium concentration is about 15\% per 0.05pH and 0.2\% per $^\circ$C \cite{Bers1994}.
To obtain a final concentration of 40nM of the CG5N indicator in the calibration solutions, we prepared a 1mM stock solution of the CG5N hexapotassium salt (Molecular Probes, C3737) in phosphate buffered saline (PBS, cp. supplementary material). With a micropipette, we added 50 $\mu$l of this stock solution to 125 $\mu$l to the calcium buffer solutions in sealable cuvettes and re\-corded the corresponding CG5N flu\-o\-res\-cence decays at a temperature of 37$^\circ$C. 

\subsection{Tissue explant cultures}
Fresh porcine tracheae were obtained from a local slaughter house. The epithelium was carefully separated from the cartilage and cut into round pieces of 20mm diameter. The tissue samples of 1~-~2~mm thickness contained the respiratory epithelium, the basement membrane as well as underlying connective tissue. Subsequent to their microdissection the explants were washed by incubation in PBS for one hour at 37$^\circ$C. Thereafter, the explants were placed onto microporous cell culture inserts (BD FALCON; 24 mm diameter, 0.4$\mu$m pore size, PET track-etched membrane) and cultured at 37$^\circ$C, 5\% CO$_2$, a relative humidity of $>$90\% and an established ALI, i.e. with cell culture medium (cp. supplementary material) present only on the basal side of the insert. Ciliary beating was regularly checked prior to experiments by light microscopy.

\subsection{Protocols for recording excitation and emission spectra of tracheal explants}
\label{sec:protocols_tissue_insert}
The spectral flu\-o\-res\-cence characteristics for homogenized porcine tracheal explant was measured with a fluorimeter (LS-5B, Perkin-Elmer). The tracheal epithelium was washed twice for 10 minutes in two PBS-baths, homogenized mechanically for 30 seconds with an ultrathurax homogenizer (Branson) and cells were disrupted by applying ultrasonic pulses for 10 minutes with an ultrasonic homogenizer (SONOPLUS HD 2070, BANDELIN electronic GMBH \& Co. KG; setting: 30\% intervals, 60\% cycles). The homogenate was centrifuged at room temperature for 1h at 1300g. The supernatant was filtrated with membrane filters in the following order: 5$\mu$m PVDF-membrane (MILLEX-GV, Millipore S.A), 0.22$\mu$m PVDF-membrane (MILLIEX-GV, Millipore S.A.) and 0.2$\mu$m Anotop 25-mem\-brane (Whatman). 


\subsection{Loading of cells with CG5N/AM}
\label{calciumgreen_loading}
A stock solution of the calcium indicator CG5N/AM (Invitrogen, SKU\# C-3739) was prepared by dissolving 50$\mu$g CG5N/AM  in 1$\mu$l dimethyl sulfoxide (DMSO for spectroscopy, Dr. Grogg Chemie \& Co.). Adding 799$\mu$l PBS resulted in 800$\mu$l final loading solution with 46$\mu$M CG5N/AM. This 800$\mu$l loading solution was applied to the surface of each tracheal explant. After incubating the explants for 3h at 37$^\circ$C, 5\% CO$_2$ and at a relative humidity of $>$90\%, the loading solution was removed and samples were rinsed with PBS to remove the dye completely from the extracellular space. The explants were incubated submerged in PBS for at least one additional hour to facilitate the de-esterification of intracellular CG5N/AM (CG5N becomes cell impermeant and remains inside the cells) and to equilibrate the samples before starting the measurement. Finally, the tracheal explants were placed back onto cell culture inserts and cultured with medium buffered for 0.04\% CO$_2$ (cp. supplementary material).


\subsection{Time resolved calcium measurements and stimulations}
\label{sec:protocols_stimulations}

Prior each measurement the sample was rinsed with PBS-solution in order to remove excess mucus. The cleaning procedure was applied {\em in situ} in the cell exposure chamber, through the silicon tubes shown in Fig.~\ref{fig:figure_1}a. Thereafter, the sample was allowed to equilibrate for about one hour. In the measurement chamber the explants were cultured at the ALI with cell culture medium (cp. supplementary material) on the basal side of the explants only at 37$^\circ$C, 0.04\% CO$_2$ and at a relative humidity of $>$90\%.

To stimulate intracellular cal\-ci\-um changes we used ionomycin cal\-ci\-um salt (Sigma, IO634) and ATP (Fluka, 02060), both dissolved in PBS. The ATP increases intracellular calcium concentration over purinergic receptors whereas ionomycin acts as a calcium ionophore \cite{Korngreen1996}. The ATP powder was dissolved directly in PBS, while the ionomycin cal\-ci\-um salt solution was prepared from a 5mM ionomycin cal\-ci\-um salt stock solution in ethanol. Stimulation of the epithelia was achieved by flushing the previously equilibrated sample with 0.5ml of the stimulant solution at the desired concentration (ionomycin: 5$\mu$M and 50$\mu$M, ATP: 50$\mu$M and 500$\mu$M). Pure PBS buffer was used as a "blank". The solutions were delivered through the centrally located silicon tube, shown in Fig.~\ref{fig:figure_1}a.

\section{Results}

\begin{figure}
\includegraphics[width=15cm]{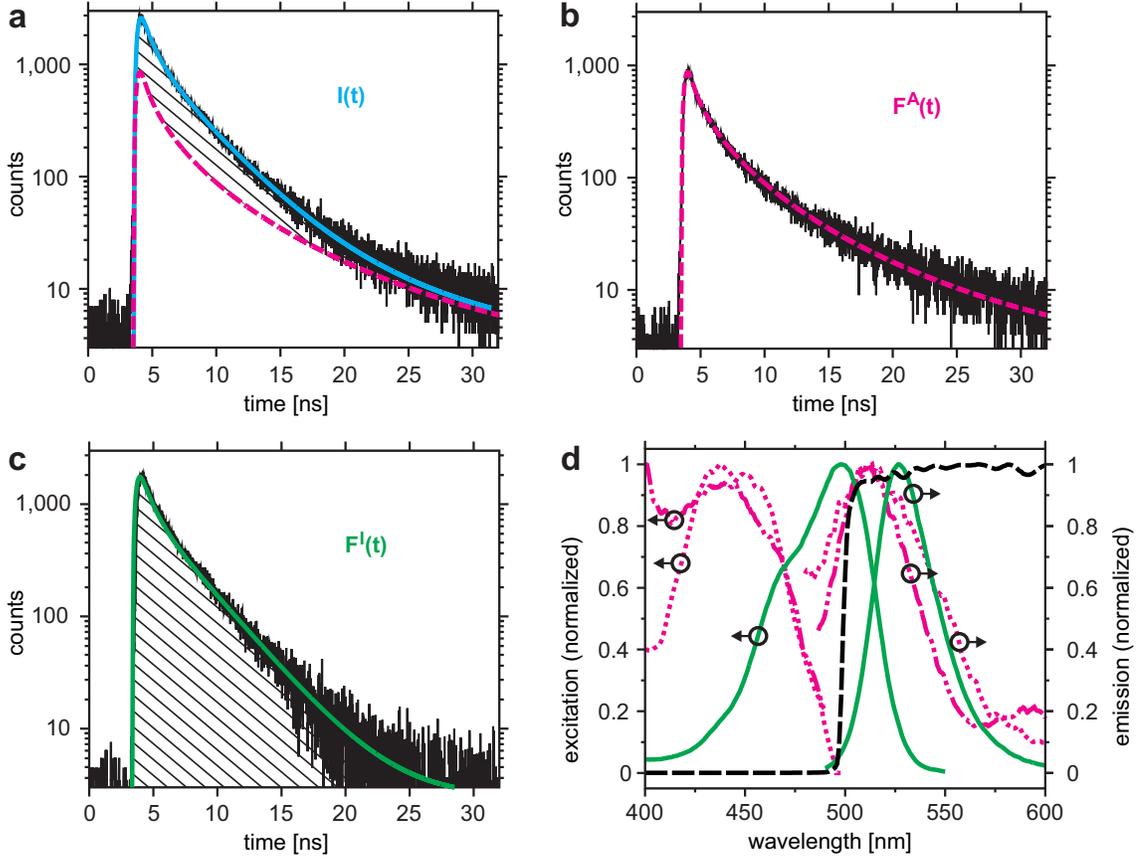}
	\caption{\label{fig:figure_2} Fluorescence of CG5N, respiratory tissue and cell culture medium and illustration of the data evaluation procedure. \textbf{(a)}~Semi-logarithmic plot of the nanosecond flu\-o\-res\-cence response from a sample loaded with the cal\-ci\-um indicator CG5N (blue -). \textbf{(b)}~Autofluorescence of the same respiratory tissue as in (a).
Fit by stretched exponential Eq.~\ref{eq:stref} (magenta~--), also shown in (a). \textbf{(c)}~Upon subtracting (b) from (a), the double exponential decay of the cal\-ci\-um indicator is visible (green~--, hatched areas in (a) and (c)). From the amplitude ratio of the fast component (initial decay) and the slow component we determine the cal\-ci\-um concentration (cp. Eq.~\ref{eq:calibration_equation}). \textbf{(d)}~The flu\-o\-res\-cence excitation and emission spectra of CG5N (green -), homogenized and filtered tracheal explant (magenta~.., prepared as described in Section \ref{sec:protocols_tissue_insert}) and of the cell culture medium (magenta -..-). The dashed black line is the transmission spectrum of the dichroic mirror used in the detection path in order to block the illuminating laser light.}
\end{figure}

\begin{figure}
	\centering
	\includegraphics [width=15cm]{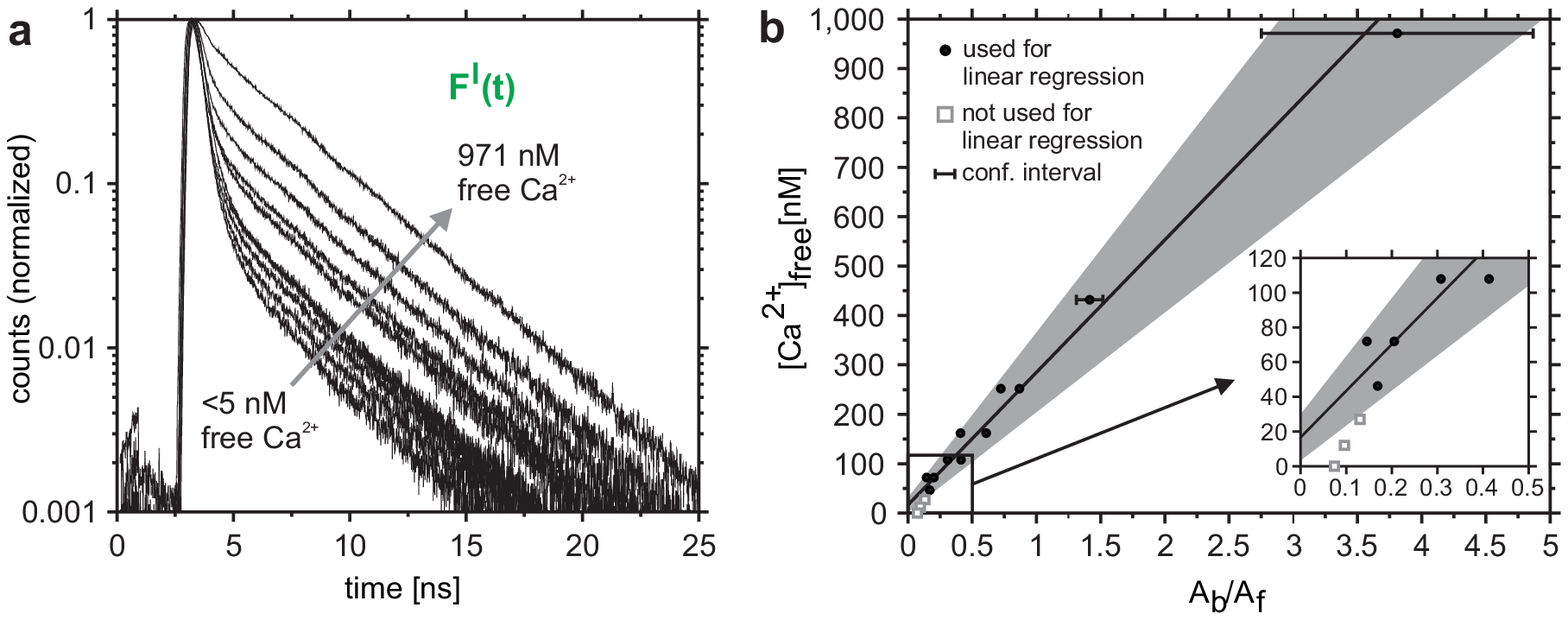}
	\caption{\label{fig:figure_3} Calibration measurements and evaluation. \textbf{(a)} Fluorescence response of the CG5N to varying free calcium concentrations in the calcium buffer solutions. pH 7.2 , 37$^\circ$C, $\left[\mathrm{Ca}^{2+}\right]$: $<$ 5, 12, 27, 46, 72, 108, 162, 252, 432, 971~nM. \textbf{(b)} Calcium concentration as a function of the ratio of the fractional amplitudes. The inset shows a magnification of the region, where non-specific ion binding to the probe becomes dominant. A $1/\mathrm{err}^2$-weighted linear regression was performed using only data points where $\left[\mathrm{Ca}^{2+}\right]>50$nM (cp.~Eq.~\ref{eq:calibration_equation_measurement}), for which we assume that non-specific ion binding can be neglected. The underlying gray area indicates the 68\%-confidence interval of the linear regression.}

\end{figure}

\subsection{Calibration}
The nanosecond responses $F^I(t)$ of CG5N recorded at varying free cal\-cium con\-cen\-trat\-ions $\left[\mathrm{Ca}^{2+}\right]_{free}$ at a temperature of 37$^\circ$C and pH 7.2 are shown in Fig.~\ref{fig:figure_3}a. The amplitudes of the two decay components, $A_{b}$ and $A_f$, show a strong dependence on $\left[\mathrm{Ca}^{2+}\right]_{free}$, but the lifetimes, $\tau_{b}$ and $\tau_f$ remain almost constant: $\hat{\tau}_f=0.49\pm 0.01$ns and $\hat{\tau}_{b}=3.36 \pm 0.01$ns. These mean values were then held fixed in a second pass of evaluation in order to recover reliable estimates of the fluorescence amplitudes $A_f$ and $A_{b}$. 
In Fig.~\ref{fig:figure_3}b the ratios $A_{b}/A_f$ are plotted versus $\left[\mathrm{Ca}^{2+}\right]_{free}$. The inset shows a zoom of the concentration range below 50nM, where the dependency becomes non-linear. The non-linearity originates from non-specific ion binding that becomes dominant at low calcium concentrations. Therefore we restrict the determination of the calibration factor $K_dR$ (Eq.~\ref{eq:calibration_equation}) to the range $\left[\mathrm{Ca}^{2+}\right]_{free}>50$nM. The weighted linear regression yielded the calibration curve
\begin{equation}
\left[\mathrm{Ca}^{2+}\right]_{free}=268\,\left(\pm 77\right)\cdot\frac{A_{b}}{A_f}+17\,\left(\pm 13\right)\,\left[\mathrm{nM}\right],
\label{eq:calibration_equation_measurement}
\end{equation}
where the values in the brackets denote the 68\%- confidence intervals. This calibration curve is shown in Fig.~\ref{fig:figure_3}b as black line and the confidence interval is indicated by the underlying gray area.



\subsection{Epithelial tissue auto\-flu\-o\-res\-cence and stray light}
\label{sec:epithelial_autofluorescence_decay}

Excitation and emission spectra of the tissue and culture medium overlap those of the indicator dye (cp. Fig.~\ref{fig:figure_2}d). Therefore the only way to separate their contributions is by analyzing the time-course of their fluorescence decay. Fig.~\ref{fig:figure_2}b shows a typical example of the nanosecond auto\-flu\-o\-res\-cence response of a tracheal explant placed on a cell culture insert and cultured at the ALI.

In order to estimate the influence of the backscattered laser light on the recovered autofluorescence mod\-el pa\-ram\-e\-ters, we performed a least-square analysis using each mod\-el for the same data: a series of 180 consecutive auto\-flu\-o\-res\-cence re\-cordings of 20s over 60 minutes was analyzed and the set of recovered pa\-ram\-e\-ters combined according to the procedure described in the supplementary material. With the mod\-el for the autofluorescence contribution $F^A(t)$, the recovered mod\-el pa\-ram\-e\-ters are $\hat{\tau}_s=0.30\pm0.02$, $\hat{h}_s=2.48\pm0.03$ and with the enhanced mod\-el $F^A(t)+S(t)$ that takes reflections into account, the recovered pa\-ram\-e\-ters are $\hat{\tau}_s=0.32\pm0.02$, $\hat{h}_s=2.47\pm0.03$. Thus, the reflections in the initial rise of the signal do not influence the recovered pa\-ram\-e\-ters $\tau_s$ and $h_s$ significantly. 

\subsection{Long term stability of epithelial tissue autofluorescence}
\begin{figure}
	\centering
	\includegraphics[width=15cm]{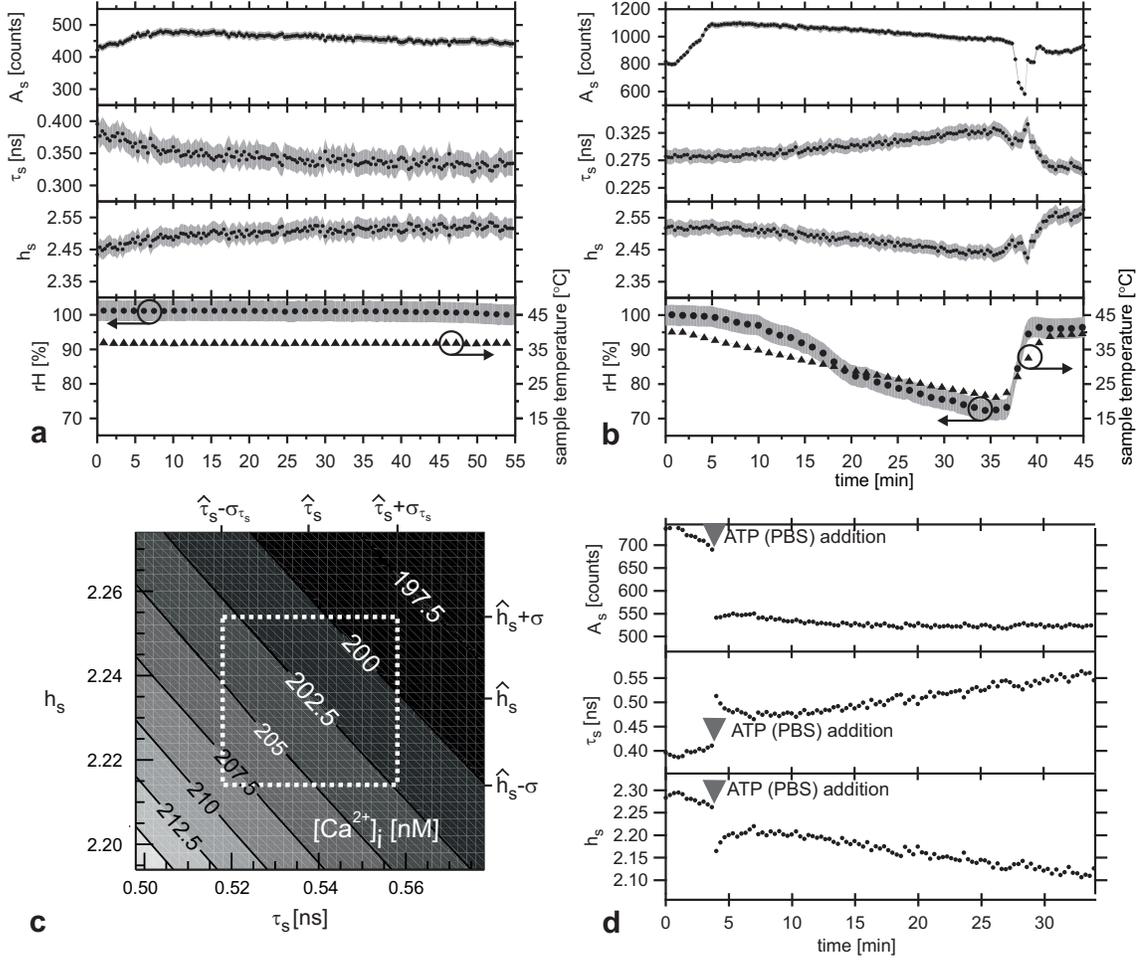}
		\caption{\label{fig:figure_4} Characteristics of the autofluorescence of respiratory tissue and culture medium in terms of the parameters of the stretched exponential function and estimation of the error in the cytosolic calcium concentration measurement introduced by fluctuations of the autofluorescence. \textbf{(a)}~Tissue autofluorescence at constant temperature and humidity (37$^\circ$C, close to 100\%rH). \textbf{(b)}~Tissue autofluorescence at temperature and humidity variation. \textbf{(c)}~Error of the calculated cal\-ci\-um concentration in dependence of the fluctuations of the stretched exponential pa\-ram\-e\-ters $\tau_s$ and $h_s$ at constant temperature and humidity. Gray scale codes the apparent $\left[\mathrm{Ca}^{2+}\right]_i$ between 195nM and 215nM. Dashed rectangle: 2-$\sigma$ area of the observed fluctuations of $\tau_s$ and $h_s$ ($\sigma_{\tau_s}=0.02$ns and $\sigma_{h_s}=0.02$). The midpoint of the map, i.e. $\hat{\tau}_s=0.54$ns and $\hat{h}_s=2.23$, corresponds to the values obtained by averaging the measurement over one hour.  \textbf{(d)}~Tissue autofluorescence response to addition of 0.5ml 500$\mu$M ATP (PBS). Immediately after addition of the ATP-solution the autofluorescence parameters change by $\Delta{h_s}=-0.1$ and $\Delta{\tau}_s=+0.1$ns that corresponds to an error in the calcium concentration measurement of 10nM.}
	
\end{figure}

Since we are dealing with living tissue, we must anticipate that autofluorescence may change with time. Therefore we re\-corded auto\-flu\-o\-res\-cence decay curves consecutively over a time period of 60 minutes in time intervals of 20 seconds. Temperature and humidity were kept constant at 37$^\circ$C and $>95$\% relative humidity. Fig.~\ref{fig:figure_4}a shows the pa\-ram\-e\-ters recovered from a series of measurements ('trachea 3' in Tab.~\ref{tab:pig_trachea_autofluorescence_consttemp}).

The amplitude of the autofluorescence contribution depends on the properties of the optical setup, such as alignment and laser power settings, and may show some drift over the observed time span. Therefore the pa\-ram\-e\-ters of interest are the form-pa\-ram\-e\-ters $\tau_s$ and $h_s$ of the stretched exponential model $F^A(t)$ (Eq.~\ref{eq:stref}). Those exhibit a drift during the first few minutes of the experiment, but then stabilize within the estimation error (1-sigma confidence interval) indicated by the underlying gray band. 
We repeated this measurement with four samples from different animals. The data summarized in Tab.~\ref{tab:pig_trachea_autofluorescence_consttemp} indicates a significant variability. Therefore the autofluorescence response must be measured for each individual sample prior to loading with indicator dye. 

\begin{table}

    \begin{tabular}{l c c} \hline \hline 
    
    sample        &   $\hat{\tau}_s\pm\sigma_{\tau_s}$ {[ns]}        &   $\hat{h}_s\pm\sigma_{h_s}$   \\ \hline 
    trachea 1  \hspace{1mm}     &   \hspace{1mm} $0.52\pm 0.02$  \hspace{1mm} &  \hspace{1mm} $2.14\pm 0.02$ \hspace{1mm} \\ 
    trachea 2      &   $0.54\pm 0.02$ &  $2.23\pm 0.02$   \\
    trachea 3      &  $0.35\pm 0.02$  & $2.50\pm 0.03$  \\
     trachea 4      & $0.34\pm 0.06$  & $2.64\pm 0.08$   \\ \hline \hline
    \end{tabular}
    
   \caption{ \label{tab:pig_trachea_autofluorescence_consttemp} Averaged autofluorescence model pa\-ram\-e\-ters $\hat{\tau}_s$ and $\hat{h}_s$ of different tracheae. For each sample a series of 180 consecutive auto\-flu\-o\-res\-cence re\-cordings of 20 seconds over 60 minutes was taken at a constant temperature of 37$^\circ$C and $>95$\% relative humidity.} 
\end{table}

\subsection{Influence of temperature and humidity on the epithelial tissue autofluorescence \\}
\label{sec:inf_of_temp_and_hum}
To assess the stability of the autofluorescence we varied the temperature while re\-cording the auto\-flu\-o\-res\-cence decay consecutively over a time period of 60 minutes in time intervals of 20 seconds. The result of this measurement is shown in Fig.~\ref{fig:figure_4}b. 
The temperature was first linearly decreased from 40$^\circ$C to 21$^\circ$C over a time span of about 35 minutes and then rapidly reset to 40$^\circ$C within about 3 minutes. Changes of relative air humidity are correlated with the temperature. All pa\-ram\-e\-ters are correlated with the variation of temperature and humidity, especially in the region of the temperature jump at t=37-40 min. The amplitude of the stretched exponential function exhibits the strongest response with a variation of about 30\% of its magnitude. The form-parameters, $\tau_s$ and $h_s$, are fortunately less sensitive: they exhibit variations of about 18\% and 5\%, respectively. Nevertheless, temperature and humidity were kept at constant values while performing autofluorescence and calcium measurements; influences of temperature and humidity variations on the calcium measurements can therefore be neglected.

\subsection{Influence of applying stimuli in form of a solution on the tissue autofluorescence \\}
An issue, which must be considered, is the change of autofluorescence in response to biochemical stimuli. Since we add solutions to evoke a change in cytosolic calcium, we have to estimate the effect of the stimuli on the epithelial autofluorescence. The disturbance is of chemical and partly also of mechanical nature, due to the flushing process. To estimate the effect on autofluorescence, we recorded the autofluorescence of a sample that was free of fluorescence indicator and stimulated it by flushing it with 500$\mu$M ATP. The time course of the autofluorescence response to this stimuli in terms of the parameters of the stretched exponential function is depicted in Fig.~\ref{fig:figure_4}d. Immediately after adding the stimuli solution the autofluorescence parameters change by $\Delta{h_s}=-0.1$ and $\Delta{\tau}_s=+0.1$ns which results in an error in the calcium concentration measurement of 10nM. These biochemical stimuli induced changes of the autofluorescence background limit the accuracy of this method for solution stimulated Calcium-measurements.

\subsection{Sensitivity of the cal\-ci\-um measurement to the epithelial tissue autofluorescence model pa\-ram\-e\-ters for the unstimulated case}
\label{sec:sens_of_calcium_det}

Since slight changes of the autofluorescence parameters appear to be inevitable, we must estimate the sensitivity of the calculated cal\-ci\-um concentration to their variations. Using for example the data from 'trachea 2' (cp. Tab.~\ref{tab:pig_trachea_autofluorescence_consttemp}), we calculated the cal\-ci\-um concentration for various fixed values of $\tau_s$ and $h_s$. Fig.~\ref{fig:figure_4}c shows the resulting apparent $\left[\mathrm{Ca}^{2+}\right]_i$ as a map with the horizontal coordinate $\tau_s$ and the vertical coordinate $h_s$. The dashed rectangle denotes the 2-$\sigma$ range of the parameter fluctuations observed in the sample 'trachea 2'. The midpoint at $\hat{\tau}_s=0.54$ns and $\hat{h}_s=2.23$ corresponds to the values obtained by averaging the measurement over one hour. Inspecting the map, we may conclude that fluctuations of the autofluorescence parameters in the order of their standard deviations, $\sigma_{\tau_s}$ and $\sigma_{h_s}$, can introduce an error in the cal\-ci\-um determination of up to about 5\%.

	

\subsection{Calcium monitoring}
\label{sec:undisturbed_sample}

To assess the stability and accuracy of the calcium measurement (cp. Fig.~\ref{fig:figure_2}) we re\-corded the response of a trachea sample loaded with CG5N at a sampling rate of 0.5Hz for 120 minutes. Using the two-pass data evaluation method discussed in the supplementary material we first determined the fluorescence life times of CG5N. The lifetimes of the intracellular CG5N differ significantly from {\em in vitro} measurements at pH 7.2.  While the change of the life time $\tau_{b}$ of the calcium bound form is negligible (3.42$\pm 0.01$ns instead of 3.36$\pm 0.01$ns), the lifetime $\tau_{f}$ of the calcium free form increases to $0.76\pm 0.03$ns, to be compared to $0.49\pm 0.01$ns measured {\em in vitro}. According to Kuhn \cite{Kuhn1993}, the short lifetime of the calcium free form can be attributed to strong quenching by an electron transfer from the calcium chelator group (BAPTA). The transfer mechanism is likely to be influenced by the spatial arrangement of BAPTA and the fluorophore \cite{Schoutteten1999} and it is conceivable that this arrangement depends on the intracellular environment. This effect would deserve a deeper investigation. On one hand, it may turn out to be useful for refined intracellular sensing, on the other hand it may represent a limitation of the calibration procedure.  However, in the case of strong quenching, fluorescence lifetime is dominated by the non-radiative decay rate, so that $q_f \propto \tau_f$. An inspection of Eq.~\ref{eq:calibration_equation} shows that in this case the effect cancels. The dissociation constant of CG5N is not sensitive to changes of the chemical environment around physiological levels \cite{Eberhard1991}, therefore we do not expect the lifetime change to cause serious errors in the determination of the calcium concentration. 

\begin{figure}
	\centering
	\includegraphics[width=15cm]{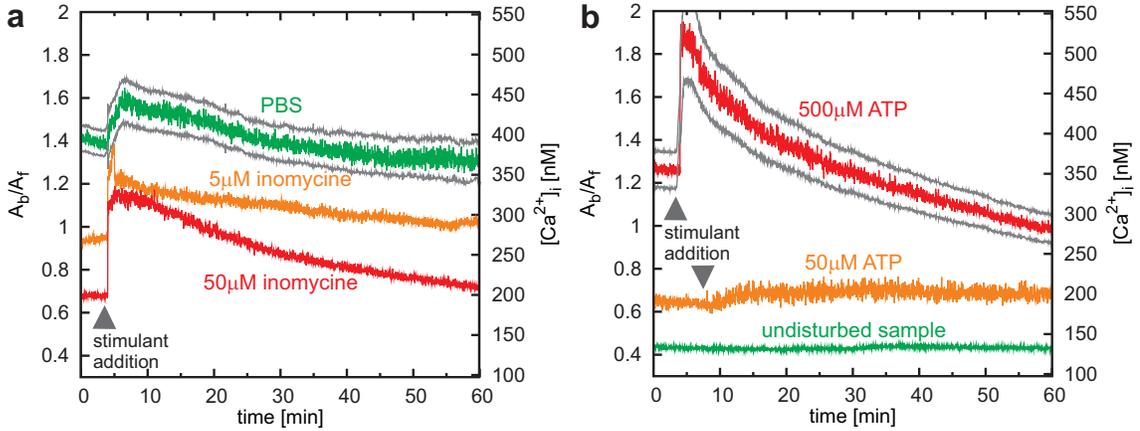}
	\caption{\label{fig:figure_5} Examples of responses of cytosolic cal\-ci\-um to stimulants. \textbf{(a)} Effect of ionomycine. \textbf{(b)} Effect of ATP. The data originate from different trachea samples. Note the variability of the baseline. The two gray lines indicate the 1-$\sigma$ error of the calcium concentration.}
\end{figure}

The time course of the recovered amplitude ratios and the corresponding cytosolic cal\-ci\-um concentration from an undisturbed tissue is included in Fig.~\ref{fig:figure_5}b. In this sample the relative contributions to the steady-state fluorescence signal are partitioned as follows: $\approx$20\% free indicator $P_f$ fluorescence, $\approx$60\% bound indicator $P_b$ fluorescence and $\approx$20\% autofluorescence.

The other time courses in Fig.~\ref{fig:figure_5} demonstrate the effect of stimulants. The tissue inside the exposure chamber was flushed with 0.5ml of the stimulant solutions (see Section \ref{sec:protocols_stimulations}). Simultaneously, the nanosecond flu\-o\-res\-cence response was re\-corded with a time resolution of 2s.

Stimulating the tissue with phosphate buffered saline (in PBS) solution containing no additional chemical stimulant can cause a rise of intracellular calcium concentration of up to $\Delta\left[\mathrm{Ca}^{2+}\right]\approx40$nM (cp. Fig.~\ref{fig:figure_5}a). We assume that this response is mainly due to mechanical perturbations of the cells \cite{sanderson1990} during the flushing process through the centrally located silicon tube (cp. Fig.~\ref{fig:figure_1}a). 

The exposure of the sample to 0.5ml ionomycin cal\-ci\-um salt solution (cp. Fig.~\ref{fig:figure_5}a) causes an immediate rise of the intracellular cal\-ci\-um concentration of $\Delta\left[\mathrm{Ca}^{2+}\right]\approx120$nM. The lower ionomycine concentration (5$\mu$M) evokes a calcium spike of few second of duration, followed by a slow decline towards the base level. The high concentration of 50$\mu$M has longer lasting effect, but the calcium also returns to the base level within about 40 minutes.

The response of the sample to the exposure to 0.5ml 50$\mu$M ATP is hardly detectable whereas the exposure to 0.5ml of 500$\mu$M ATP causes a strong rise of the intracellular cal\-ci\-um concentration of ($\Delta\left[\mathrm{Ca}^{2+}\right]\approx200$nM) that is followed by a slow decrease. However, we often observe an undershoot, i.e. the intracellular calcium concentration falls below and does not return to the original value. 

The large variability of the calcium baseline is characteristic for native, non-standardized tissue samples, but the observed levels between 120 and 350nM are unusually high.

\section{Discussion}

We demonstrated the feasibility of monitoring the state of tissue cultures {\em in situ} during the culturing process and in the presence of strong autofluorescent background. The fluorescence-lifetime-based technique allows unambiguous discrimination of the autofluorescence alone on the grounds of its nanosecond signature. Despite the simplicity of the fiber-optical setup, the sensitivity of the photon counting is sufficient to achieve a monitoring resolution of less than one second, without disturbing the sample by strong irradiation. The simple fiber-optic hardware could easily be incorporated in the tight space of the cell culture chamber. This simple fiber-optic hardware might in future also be adapted for endoscopic {\em in vivo} measurements.

After calibration absolute calcium concentrations can be measured. It appears, that there are reasonable intervals in the lower range of the spectrum (below 500nM free calcium), but in higher concentrations, the confidence interval gets significantly larger and the measurements are less accurate (Fig.~\ref{fig:figure_3}b). However, the method is more precise in the lower concentration range which is the biological important range.

To proof the principle, we employed the technique for monitoring the cytosolic calcium levels in porcine tracheal explants
cultured at the ALI. These explants include not only the epithelium but also some of the underlying tissue structures, and therefore the autofluorescence background is particularly strong.
Nevertheless, we are able to detect changes of the intracellular calcium level as small as 2.5nM for undisturbed tissues and 10nM for tissues that were biochemically stimulated. Biochemical stimulation induces a change to the autofluorescence that limits the resolution.

The response to ATP shown in Fig.~\ref{fig:figure_5}b is somewhat unexpected. Applying 0.5ml of 50$\mu$M ATP solution, a concentration that is often employed in comparable stimulation experiments \cite{korngreen1994}, we detect hardly any calcium response, in contrast to Tarran et. al \cite{tarran2005}, who found about 200nM increase in calcium concentration with 100nM ATP in cell cultures. The difference may be explained by the different techniques used: (i) the cell cultures used by Tarran et. al consist of the epithelium only, which forms a dense cell layer that tightly separates the upper from the lower compartment. This is not the case with our explant cultures, which can be compared with a (special kind of) biopsy that is placed on the transwell membrane. So, the upper and lower compartments are not tightly separated from each other. (ii) Tarran et. al perfused the epithelial surface and measured the respective ions in the fluid collected over time. 

The observed calcium baselines in the range between 120-350nM are considerably higher than the values usually measured in isolated tracheal cells of less than 100nM \cite{Salathe95,Qu1995,Evans1999,Kanoh99}, but comparable with those reported for rabbit tracheal epithelium \cite{korngreen1994}. One possible reason for the high cytosolic calcium levels is that the tissue responds to manipulations by releasing ATP. However, another plausible reason is the high glucose content of the culture medium (25mM), that also contains CaCl. Intracellular calcium elevation by D-glucose up to the levels observed in the present study has been demonstrated for several cell types \cite{Barbagallo-glucose,Song-high-glucose,Li2007-viability}. Moreover, glucose interferes in a complicated fashion with ATP related processes \cite{Gilbert-ATP-calcium}. Thus, glucose phosphorylation is the likely reason for the observed lack of response to low concentrations of ATP and for the unexpected undershoot following stimulation by 500$\mu$M of ATP. High glucose, accompanied by high calcium levels, has been shown to impair the cell viability \cite{Li2007-viability,Miwa2003}. Thus, the present feasibility study already demonstrates the usefulness of {\em in situ} monitoring: we must conclude that high glucose DMEM should be used only if absolutely necessary. However, our explant cultures needed the high-glucose medium -- cultivation with low-glucose medium would have imposed stress on the cells.

\section{Conclusion}
We have demonstrated in-real time monitoring of cell and tissue activity in presence of strong autofluorescence background with a time resolution of less than 1s at the example of {\em in situ} cytosolic calcium monitoring. The low affinity fluorescence indicator CG5N was introduced to the cells. Fluorescence-lifetime-based technique allows unambiguous discrimination of the autofluorescence from the CG5N fluorescence alone on the grounds of their nanosecond signature. From the nanosecond signature of CG5N the calcium concentration can be deduced.

In biochemically stimulated tissue explants the stimuli induced a significant change of tissue-autofluorescence. Therefore we conclude that for a measurement where a stimuli is added as a solution, the measurement accuracy of this method for Calcium-measurements is limited by this biochemical stimuli induced change of the autofluorescence background. In our specific case the detection limit of cytosolic calcium changes was 10nM.

In unstimulated tissue explants the accuracy of measuring cytosolic calcium levels is limited to 2.5nM due to the fluctuations of tissue-autofluorescence.

Many questions are still open, such as the high calcium levels, the low response to ATP and the large variability of the baselines. In future work those issues should be addressed using standardized tissue samples.

\begin{figure}
	\centering
	\includegraphics[width=10cm]{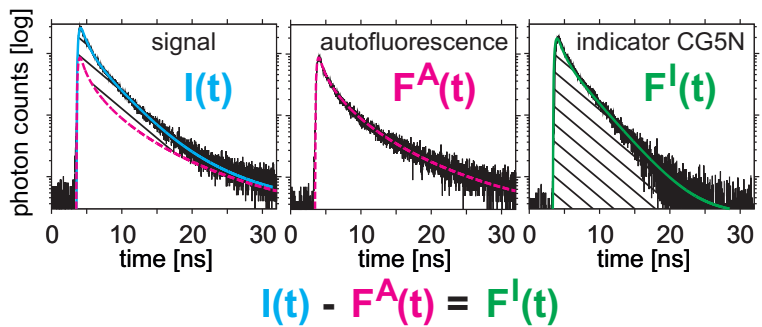}
	\caption{\label{fig:abstract_figure} ``Abstract Figure'': Illustration of measurement-principle. The semi-logarithmic plots depict from left to right the fluorescence response $I(t)$ of the tissue explant culture loaded with CG5N, the autofluorescence $F^A(t)$ and the calculated calcium indicator CG5N $F^I(t)$ from which the absolute cytosolic calcium concentration is determined.}
\end{figure}

\subsection*{Acknowledgments}
We thank R. Nyffenegger and B. Kupferschmid for their assistance in developing and optimizing the protocols and solutions, and R. Friedrich for his support in developing and building the experimental setup. L. K\"{u}nzi acknowledges the support by Swiss National Science Foundation Grant K-32K1-120524. M. Ryser acknowledges the partial support from EC project POLYSOA.

\section*{Supplementary Material: Iterative reconvolution fitting (two pass procedure)}
The parameters $\{p\}$ of the model impulse response $I(t;\{p\})$ were fitted to the measured decay profiles $M\left(t \right)$ by iterative reconvolution. Thereby one assumes that
\begin{equation}
M(t;\{p\})=I(t;\{p\})*R(t)+b+\epsilon(t),
\label{eq:model_function}
\end{equation}
where $R(t)$ is the instrumental response, $b$ is a constant background due to detector dark counts or ambient light (usually negligible) and $\epsilon(t)$ the counting error. The aim is to minimize the chi-square value $\chi^2$. In the case of photon counting $\chi^2$ is estimated as \cite{becker2005}
\begin{equation}
\chi^2=\sum_{t=1}^{k}{\frac{\epsilon(t)^2}{M(t)}}.
\label{eq:chisquare}
\end{equation}
The minimization is achieved by the standard \textit{Levenberg--Marquardt} technique and quality of the fit is judged by plotting the normalized deviations $\sqrt{\epsilon^2/M}$.
The confidence intervals of the best fit parameters were estimated by the so-called exhaustive search procedure \cite{johnson1983,Roelofs1992}. Intervals obtained in such a manner take the correlations between the pa\-ram\-e\-ters into account, giving thus a more realistic estimate of the accuracy than the commonly used standard error \cite{Roelofs1992}.

The primary data set of a typical time resolved measurement of the changes of calcium concentration consists in about 2000 decay curves, 4096 time channels each. The individual records, obtained within about 2s each, are rather noisy. Therefore, for an accurate determination of the amplitudes $A_f$, $A_{b}$ and $A_s$ (Eqs.~\ref{eq:bi-exponential-impulse-response} and \ref{eq:stref}) it is important to exploit the whole data set. The indicator decay times $\tau_f$ and $\tau_b$ can be assumed to be the same for all partial measurements. This also applies for the afore measured autofluorescence parameters $\tau_s$ and $h_s$. Thus, only the amplitudes $A_f$, $A_{b}$ and $A_s$ are expected to vary. Rigorously one should simultaneously evaluate all 2000 measurements, but we employ a simple two pass procedure: in a first pass we fit the mod\-el-pa\-ram\-e\-ters $A_s$, $A_f$, $\tau_f$, $A_{b}$ and $\tau_{b}$. (Autofluorescence parameters $\tau_s$ and $h_s$ are held fixed to the previously determined values.) In a second pass we fix the parameters $\tau_f$ and $\tau_b$ at their average-values and obtain the final estimates of the amplitudes.

\section*{Supplementary Material: Cell culture media and PBS}
Depending on wheth\-er the cell cultures were maintained in an incubator with 5\% CO$_2$, or in the exposure chamber during the flu\-o\-res\-cence measurements with only 0.04\% CO$_2$ (ambient air), two differently buffered cell culture media were used.

\subsection*{Cell culture medium for 5\% CO$_2$} 
As culture medium at standard incubator conditions we used DMEM (Invitrogen, SKU\# 31966-021, 1X, liquid, high glucose) to which we added 10\% fetal calf serum (FCS, Amimed), 2.5$\mu$g/ml amphotericin (Invitrogen, SKU\# 15290-026), 50$\mu$g/ml gentamicin (Invitrogen, SKU\# 15750-037), 10U/ml penicillin and 10$\mu$g/ml stre\-pto\-mycin (Invitrogen, SKU\# 15140-122). Finally, the pH was adjusted to 7.2 at 37$^\circ$C by adding 2N NaOH or HCl.

\subsection*{Cell culture medium for 0.04\% CO$_2$} 
Usually, cell culture medium is buffered with sodium-bicarbonate buffer to obtain a physiological pH at 5\% CO$_2$. At ambient air condition the high sodium-bicarbonate buffer concentration is not in equilibrium with the ambient CO$_2$ concentration which leads to a long term shift of the pH. Therefore we used HEPES-buffer (Fluka) instead. We dissolved 6.74g of the DMEM without sodium-bicarbonat (Invitrogen, SKU\# 12800-116, powder, high glucose) in H$_2$O Milli-Q (Millipore) and added 25mM HEPES. FCS, amphotericin, gentamicin and pen\-i\-cil\-lin-streptomycin were added at the same concentrations as listed for the medium at 5\% CO$_2$. Finally the pH was adjusted to 7.2 at 37$^\circ$C by adding 2N NaOH or HCl.

\subsection*{Phosphate buffered saline (PBS)} 
The buffer was prepared by mixing two solutions: Solution 1 contained 0.2g KCl, 8g NaCl, 2.31g Na$_2$HPO$_4$ (decahydrate), 0.2 g KH$_2$PO$_4$ (anhydrous) and 990ml H$_2$O Milli-Q; Solution 2 contained 0.133g CaCl (dihydrate), 0.1g MgCl (hexahydrate) and 10ml H$_2$O Milli-Q. Solution 2 was then added very slowly to solution 1 under vigorous stirring. After mixing the pH was adjusted to 7.2 at 37$^\circ$C by adding 2N NaOH or HCl. This procedure prevents the precipitation of calcium- and mag\-ne\-si\-um\-phos\-phate.


\bibliographystyle{elsarticle-num} 
\bibliography{calcium_paper}

\end{document}